\documentclass [showpacs, showkeys, amsmath, twocolumn, amssymb, prl, floatfix]{revtex4}
\usepackage{graphicx}
\usepackage{epsfig}
\usepackage{dcolumn}
\usepackage{bm}
\usepackage{color}

\begin{document}

\title{Hyperbolic statistics and  entropy loss in the description of gene distributions}
\author{K.~Lukierska-Walasek }
\email{k.lukierska@uksw.edu.pl}
\affiliation {Faculty of Mathematics and Natural Sciences, Cardinal Stefan Wyszy\' nski University, Dewajtis 5, 01-815 Warsaw, Poland}
\author{K.~Topolski}
\email{topolski@math.uni.wroc.pl}
\affiliation {Institute of Mathematics, Wroclaw University, Pl.  Grunwaldzki 2/4, 50-384 Wroclaw, Poland}
\author{K.~Trojanowski}
\email{k.trojanowski@uksw.edu.pl}
\affiliation {Institute of Computer Science, Polish Academy of Science, ul.Jana Kazmierza 5,01-248 Warszawa,Poland}

\begin{abstract}
 Zipf's law  implies the statistical distributions of hyperbolic type, which can describe the properties of stability and entropy loss in linguistics. We present the information theory from which follows that if the system is described by distributions of hyperbolic type  it leads to the possibility of entropy loss. We present the number of repetitions of genes in tne genomes  for  some bacteria, as {\em Borelia burgdorferi} and {\em Escherichia coli}.
 Distributions of repetitions of genes in genome appears to be represented by distributions of hyperbolic type.
\end{abstract}

\pacs{ }

\maketitle

%Zipf's law is an empirical observation which relates rank and frequency of %words in natural language \cite{1}. The law suggests modeling by %distributions of hyperbolic type \cite{2}.  We consider a class of %distributions $P= (p_{1},p_{2},. ..)$ over $N$.  If $ p_{1}\ge,p_{2}\ge %$... , $P$ is said to be hyperbolic if for given $\alpha>1$   $p_{i} \ge %i^{-\alpha}$ for all indexes $i$.   As an example we choose $ p_{i}\sim %i^{-1}( \lg(i) )^{-c}$ for some $c>2$. \\
%\section{Introduction}
{\small{\em  1. Introduction.}}
In last years   appeared the possibility to provide some knowledge of the genome sequence data in many organisms . The genome have been studied intensively by number of different methods \cite{1}-\cite{10}. The statistical analysis of  DNA is complicated because of its complex structure; it consists of coding and non coding regions, repetitive elements, etc., which have an influence on the local sequence decomposition.
The long range correlations in sequence compositions of DNA
is much more complex than simple power law; moreover effective exponents of scale regions are varying considerably between different species \cite{9}, \cite{10}. In  papers \cite{5}, \cite{6}  the Zipf approach to analyzing linguistic texts has been extended to statistical study of DNA base pair sequences. \\
In our paper we take into account some linguistic features of genomes to study the statistics of gene lengths. We present the information theory from which follows that if the system is described by special distribution of hyperbolic type it implies a possibility of entropy loss.
   Distributions of hyperbolic type describe also property of stability and flexibility which explain that the language considered in the paper \cite{11} can develop without changing its basic structure.  Similar situation can occur in  a genome sequence data which carries the genetic information. We can expect above features in gene strings as in language strings (words, sentences) because of presence of redundancy in both cases.\\
In $Sect.2$ we shall present some common features of linguistics and genetics. In $ Sect.3$ we describe Code Length Game, Maximal Entropy Principle and the theorem about hyperbolic type distributions and entropy lost. Final $Sect.4$ contains some applications to genetics and final remarks. . The  distributions appear to be hyperbolic distributions distributions in sense of theorem.\\
{\small {\em 2. Some common features of linguistics and genetics.}}
 A language is characterized by some alphabet with letters: a, b, c, ...., which form words as sequence of $ n$ letters.  In quantum physics the analogy to letters can be attached to pure states, and  texts correspond to mixed general states.  Languages have very different alphabets: computers 0,1 (two bits), English language 27 letters with space and  DNA four nitric bases: G({\em guanine}), A({\em adenine}), C({\em cytosine}), T({\em thymine}). The collection of letters can be ordered or disordered.
 To quantify the disorder of different collections of the letters we use an entropy. % With an alphabet of n letters it is possible to write a considerable number of  $N = n^m $ different texts containing m letters. %Supposing that all letters are occur with the some probability then the total number of states is $N = n^m $.
% If all states are equiprobable, each text with m letters has probability $ p_{i} = 1/n^{m}$ and % the Shannon entropy of such system is:
\begin{equation}
 \label{eq1}
  H = - \sum\limits_{i=1} p_{i}\log_{2}p_{i},
\end{equation}
where $p_{i}$ denotes probability of occurrence i-th letter.
If we take base of logarithm $2$ this will lead to the entropy measured in bits. When all letters have the same probability in all states obviously the entropy has maximum value $H_{max}$.\\
A real entropy has lower value $ H_{eff}$, because in a real languages the letters have not the same probability of appearance. Redundancy $R$ of language is defined \cite{11} as follows
\begin{equation}
 \label{eq2}
  R = \frac{H_{max} - H_{eff}}{H_{max}}
\end{equation}
The  quantity $H_{max}$\, - $H_{eff} $ is called an information.
 Information depend on difference between the maximum entropy and the actual entropy. The bigger actual entropy  means the smaller redundancy.  Redundancy can be measured by values of  the frequencies with which different letters occur in one or more texts.
Redundancy $R $ denotes the number that if we remove the part $R$ of the
letters determined by redundancy, the meaning of the text will be still understood. In English some letters occur more frequently than other and
similarly  in DNA  of vertebrates the frequency of nitric bases C and G pairs is usually less frequent than A and T pairs.  The low value of redundancy allows in  easier way to fight transcription errors in gene code. The papers \cite{5}, \cite{6} it is  demonstrated that non coding regions of eukaryotes display a smaller entropy and larger redundancy than coding regions.\\
{\small{\em 3.  Maximal  Entropy  Principle and Code Length Game}}.
In this section we shall provide some mathematics from the information theory which will be helpful in the quantitative formulation of our approach, for details see \cite{11}.\\
 Let $A$ be the $alphabet$ which is a discrete set finite or countable infinite.
 Let $M_{+}^{1}$ and $^{\sim}M_{+}^{1}(A)$ are respectively, the set of probability measures on $A$ and the set of non-negative measures $P$, such that $P(A)\leq 1$.
The elements in $A$ can be thought as {\em letters}.
%Measures in $M_{+}^{1}$  are probability distributions P(A) for which $ P(A) = 1 $ and measures in $^{\sim}M_{+}^{1}(A)$ are  non-probability distributions P(A) for which $P(A)\le 1$.
By K(A) we denote the set of mappings,  {\em  compact codes}, $k\,:A \,\rightarrow[0 , \infty]$, which satisfy {\em Kraft's equality} \cite{13}
\begin{equation}
 \label{eq3}
  \sum\limits_{i\in A}\exp (-k_{i}) = 1.
\end{equation}
 By $^{\sim}K(A)$ we denote the set of all mappings, {\em general codes}, $k\,: \, A\rightarrow[0 ,\infty]$, which satisfy $ Kraft's  inequality $ \cite{13}
\begin{equation}
 \label{eq4}
  \sum\limits_{i\in A}\exp (-k_{i}) \le 1.
\end{equation}
For $ k\in ^{\sim}K(A)$ and $i\in A$,  $k_{i}$ is the {\em code lenght}, for example the length of the word.
 For $ k\in$ $^{\sim}K(A)$ and $P \in M_{+}^{1}(A)$ the {\em average code length}  is defined as
\begin{equation}
 \label{eq5}
 <k,P> =\sum\limits_ {i\in A} k_{i}p_{i}.
\end{equation}
%\\for $ k\in$ $^{\sim}K(A)$ and $P \in M_{+}^{1}(A)$.\\
There is bijective correspondence between $p_{i}$ and $ k_{i}$
$$ k_{i} = -\ln p_{i}   \quad \mbox{and} \quad p_{i}=\exp(-k_{i}).$$
% If Kraft's inequality \cite{3} is satisfied than $k_{i}\ge-\ln p_{i}$.
For $P \in M_{+}^{1}(A)$ we also introduce the entropy
\begin{equation}
 \label{eq6}
 H(p) = -\sum\limits_{i\in A} p_{i}\ln p_{i}.
\end{equation}
The entropy can be represented as minimal average code length, (see \cite{11}):
$$H(P)=\min_{k\in ^{\sim}K(A)}<k,P>.$$
Let ${\cal P}  \subseteq{ \cal M }_{+}^{1}(A)$  than
\begin{eqnarray}
\label{test}
 H_{\max}(\cal P)&=&\sup_{P\in \cal P} \inf_{k\in ^{\sim}K(A)}< k, P >\nonumber\\
&=&\sup_{P\subset \cal P}H(P)\nonumber\\
 &\leq&\inf_{k\in ^{\sim}K(A)} \sup_{P\subset \cal P}< k, P > \nonumber\\
&=& R_{min}(\cal P).
\end{eqnarray}
The formula $7$  present the Nash optimal strategies. ${\cal R}_{min}$ denotes $ minimum$ $risk$, $k$ denotes the Nash equilibrium code, $P$ denotes probability.  For example, in a linguistics the listener is a minimizer, speaker is a maximizer. We have words with distributions $p_{i}$ and their codes $k_{i}$, $i=1,2,... $. The listener chooses  codes $k_{i}$,the speaker chooses probability distributions $p_{i}$.\\
We can notice that Zipf argued \cite{14} that in the development of a language vocabulary balance is reached as a result of two opposing forces: $unification$ which tends to reduce the vocabulary and corresponds to a principle of least effort, seen from point of view of speaker and {\em diversification} connected with the listeners wish to know meaning of speech.\\This principle $ (7)$ is so basic as {\em Maximum Entropy Principle} has a sense that search for one type of optimal strategy called as {\em  Code Length  Game} translates directly into a search for distributions with maximum entropy.
It is a given a code  $ k\in\,  ^{\sim}K(A) $ and distribution $P\in M^{1}_{+}(A)$. Optimal strategy according  $ H(P) = inf_{k\in\, ^{\sim}K}<k,P>$ is represented by entropy $H(P)$, where actual strategy is represented by $<k,P>$.\\
Zipf's law is an empirical observation which relates rank and frequency of words in natural language \cite{14}. This law suggests modeling by distributions of hyperbolic type \cite{11}, because no distributions over  $N$ have probabilities proportional to $1/i$, due to the lack of normalization condition.\\ We consider a class of distributions $P= (p_{1},p_{2},. ..)$ over $N$.  If $ p_{1}\ge,p_{2}\ge ... $, $P$ is said to be hyperbolic if for any given $\alpha>1$,   $p_{i} \ge i^{-\alpha}$ for infinitely many indexes $i$.   As an example we can choose $ p_{i}\sim i^{-1}( \log(i) )^{-c}$ for some constant  $c>2$.\\
The code lenght game for model ${\cal P}\in M_{+}^{1}(N)$  with codes  $k:  A \rightarrow  [0 ,\infty]$ for which $ \sum\limits_{i\in A}\exp (-k_{i}) = 1$, is in equilibrium if and only if $ H_{max}( co ({\cal P }) = H_{max}({\cal P})$. In such a case
a distribution $P^{*}$ is the $H_{max}$ attractor such that $P_{n}\rightarrow P^{*}$, if for every sequence $ (P_{n})_{n>1} \subseteq \cal P $  for which $H( P_{n}) \rightarrow H_{max}(\cal P)$. One expects that $ H(P^{*}) = H_{max}(\cal P)$ but in the case with entropy loss we have $ H(P^{*}) < H_{max}(\cal P)$.  Such possibility appears when distributions  are hyperbolic. It follows from  theorem  in \cite{11}.\\[2mm]
{\bf Theorem.} {\em
Assume that $P^{*}\in M_{+}^{1}(N)$  is of finite entropy and it has ordered point probabilities. Than necessary and sufficient condition that $P^{*}$ can occur as $ H_{max}$- attractor in a model with entropy loss, is that $P^{*}$ is hyperbolic. If this condition is fulfilled than for every $h$ with
    $ H(P^{*})< h < \infty $, there exists a model ${\cal P} = {\cal P}_{h}$ with $P^{*}$ as    $H_{max}$--attractor and $ H_{max}({\cal P}_{h}) = h $.  In fact, ${\cal P}_{h} = ( P|<k^{*},P> \le h) $ is a largest model.  $k^{*}$ denotes the code adopted to $ P^{*}$, i. e $k^{*} = -\ln(p^{*})$, $ i>1$.}\\[2mm]
As example we can consider "an ideal" language where the frequencies of words are described by hyperbolic distribution $ P^{*} $ with finite entropy. At a certain stage of life one is able to communicate at reasonably height rate about $H(P^{*})$ and improve language by introduction of specialized words, which occur seldom in a language as a whole. This process can be continued during the rest of life. One is able to express complicated ideas develop language without changing a basic structure of the language. Similar situation we can expect in gene strings, which carry an information.\\
{\small \em{ 4. Application to genetics.}}
%The special systems where maximum entropy distributions does not exists can be  described by distributions of hyperbolic type. In such systems stability and flexibility is present similarly as in natural languages or in description of genes where redudency is confirmed.
%From theorem it follows that in such a systems as natural languages or gene %sequences maximum entropy is not reached.
The gene  data which we shall use were obtained from GeneBank (ftp://ftp.ncbi.nih.gov).
\begin{figure}[h]
\centerline{\includegraphics[width=7.1cm]{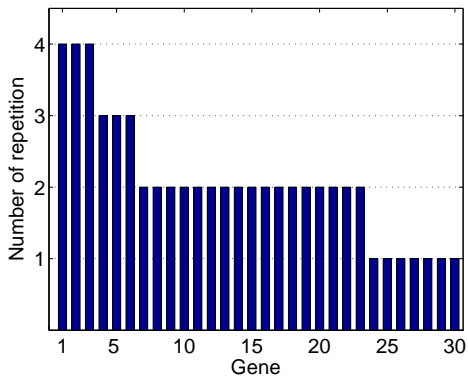}}
\caption{\label{Fig1} Graph of the number of repetition of genes in genome of the {\em Escherischia coli}. Genom consists of 5874 genes }

\end{figure}

\begin{figure}[h]
\centerline{\includegraphics[width=7.1cm]{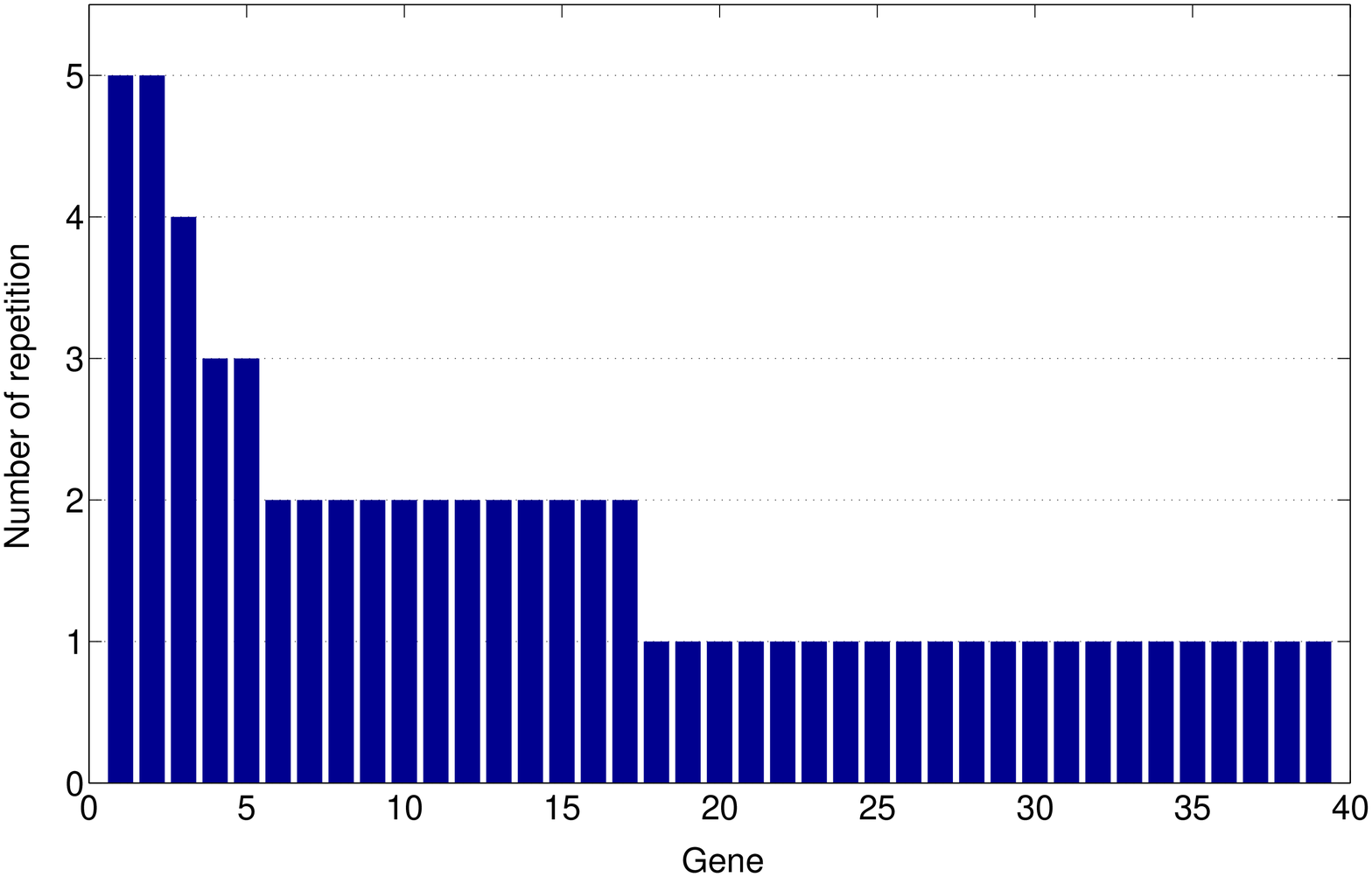}}
\caption{\label{Fig2} Graph of the number of repetition of genes in genome of the {\em Escherischia coli}. Genom consists of 4892 genes }

\end{figure}

%Figure 1 presents histogram of the gene length in genome of the  {\em Borrelia burgdorferi} and its  approximation by probability the Asymetric Inverse Gaussian Distribution %with parameters $\mu=420.9643$, $\sigma=466.0570$ and $\gamma=468.7277$.

\begin{figure}[h]
\centerline{\includegraphics[width=7.1cm]{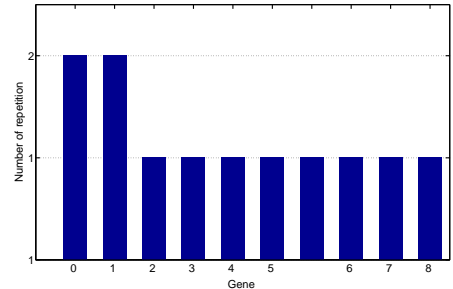}}
\caption{\label{Fig3} Graph of the number of repetition of genes in genome of the {\em Borrelia burgdorferi}The genom consists of 1239 genes}.
\end{figure}

 Figures 1,2 and 3 show how often the same genes are repeated in gemome.For example,Figure 2 shows that mostly repeated gen has number  $1$  and it appears 5 times, the second one also appears 5 times, third gene 4 times ect.The probability of  apperence of genes present hyperbolic type distributions [11].
%\begin{figure}[h]
%\centerline{\includegraphics[width=7cm]{Figg3}}
%\caption{\label{Figg3} The comparison of the tail of gene length
%distribution of the {\em Borrelia burgdorferi} to the tail of hyperbolic %type distribution.}
%\end{figure}

%\begin{figure}[h]
%\centerline{\includegraphics[width=8.6cm]{Fig3}}
%\caption{\label{Fig3} The comparison of the tail of gene length distribution of the {\em Borrelia burgdorferi} to the tail of the power type distribution.}
%\end{figure}

%On the Figure 3 we presents the fit of the tail of gene length distribution of %the {\em Borrelia burgdorferi} to function of the form $f(x)=C\,x^{-1-a}$, where %$C>0$ and $a>2$ are constants. The least squares best fit are obtained with %$C=1500,899$ and $a=2,165$.\\
%\begin{figure}[h]
%\centerline{\includegraphics[width=5.6cm]{Fig4}}
%\caption{\label{Fig4}
%\end{figure}

%\begin{figure}[h]
%\centerline{\includegraphics[width=7.3cm]{Fig5}}
%\caption{\label{Fig5} Plot of the logarithm of the gene length $L$ versus logarithm of its length rank for the {\em Saccharomyces cerevisiae S288c} genome.               }
%\end{figure}

%\begin{figure}[h]
%\centerline{\includegraphics[width=7.3cm]{Fig6}}
%\caption{\label{Fig6} Plot of the logarithm of the gene length $L$ versus logarithm of it length rank for the genes with size bigger  then 5379 bp in  {\em Saccharomyces %cerevisiae S288c} genome. The strain line represent the linear regression  line $y=-0.2413\,x+8.4718.$}
%\end{figure}

{\small {\em 5. Final remarks}.} Zipf's law relates the rank $k$ and the frequency of words $f$ in natural languages by exact hyperbolic formula $ k \times f = const $. However the probability values $1/n$ for all natural numbers $n$ do not satisfy the normalizability condition. This problem is not occurring if we introduce distributions of hyperbolic type. In the paper we follow the statistical description of the structure natural languages and we present the informatical considerations  providing as conclusion the stability property and the possibility of entropy loss. We apply similar reasoning to description of genoms with the  genes corresponding to words in languages. We test three examples of genomes and describe the gene  statistics dealing with the number of repetitions  in genome. It appears that the describution of genes is  reprezented  by dystributions of hyperbolic type.

\begin{acknowledgments}
One of the authors (K. L-W) would like to thank prof. Franco Ferrari for discussions.
\end{acknowledgments}

\bibliographystyle{apsrev}

\end{document}